# Rattling and Superconducting Properties of the Cage Compound $Ga_xV_2Al_{20}$

Zenji HIROI[1], Atsushi ONOSAKA[1], Yoshihiko OKAMOTO[1], Jun-ichi YAMAURA[1], and Hisatomo HARIMA[2]

[1]*Institute for Solid State Physics, University of Tokyo, 5-1-5 Kashiwanoha, Kashiwa, Chiba 277-8581, Japan*

[2]*Department of Physics, Kobe University, Kobe 657-8501, Japan*

Low-energy rattling modes and their effects on superconductivity are studied in the cage compound $Ga_xV_2Al_{20}$. A series of polycrystalline samples of $0 < x \leq 0.6$ are examined through resistivity, magnetic susceptibility, and heat capacity measurements. A weak-coupling BCS superconductivity is observed below $T_c$ = 1.4-1.7 K for all the samples. For small Ga contents below 0.20, approximately 30% of the cages are occupied by rattling Al atoms having an Einstein temperature $T_E$ of 23 K, probably with most Ga atoms substituting for the cage-forming Al atoms. For higher Ga contents, approximately 0.05 Ga and 0.25-0.35 Al atoms coexist statistically inside the cages and behave as rattlers with $T_E \sim$ 8 and 23 K, respectively. A significant effect of Ga rattling on the superconductivity is clearly evidenced by the observation of a sharp rise in $T_c$ by 8% at $x$ = 0.20 when 0.05 Ga atoms are introduced into the case. Probably, the electron-phonon interaction is significantly enhanced by an additional contribution to the phonon density of states from the extremely low energy rattling modes of Ga atoms. In addition, a large softening of the acoustic modes is observed for $x \geq 0.20$, suggesting that the cage itself becomes anomalously soft in the presence of low-energy Ga rattling modes.

## 1. Introduction

Rattling is a local and essentially anharmonic oscillation with an unusually large atomic excursion of an atom confined in an oversized atomic cage in crystals.[1-3] Since structural coupling between a guest atom (rattler) and the surrounding cage is too weak to generate a dispersive mode, rattling can be a local mode approximated by an Einstein mode within the harmonic approximation. As the size mismatch between a rattler and a cage increases, the first excited vibration level shifts downward to result in an unusually low-energy excitation. Such a low energy mode can strongly couple with conduction electrons and determine the low-temperature properties of metallic cage compounds. Thus, what is fascinating about rattling is the electron-rattler (e-r) interaction that causes interesting phenomena. Although rattling can be approximated as an Einstein mode within the harmonic approximation, it should be noted that anharmonicity plays a crucial role in determining the e-r interaction.[3-7]

Rattling phenomena have been extensively studied in three cage compounds: Si-Ge clathrates,[8] filled-skutterudites,[2] and β-pyrochlore oxides.[9,10] The former two compounds have attracted the attention of many researchers because rattling vibration may suppress thermal conductivity, resulting in the increase in thermoelectric efficiency.[1,2] On the other hand, it has been clearly demonstrated in β-pyrochlore oxides that a large e-r interaction is helpful for Cooper pairing to induce superconductivity and thus enhancing $T_c$ as well as inducing an extremely strong-coupling superconductivity at $T_c$ = 9.60 K in $KOs_2O_6$.[3,5-7] It also causes a large mass enhancement[11,12] and a characteristic scattering of conduction electrons that gives a concave-downward temperature dependence of resistivity at high temperatures and $T^2$ resistivity at low temperatures.[4]

We now focus on another cage compound, $A_xV_2Al_{20}$ ($Al_{10}V$). This compound crystallizes in the $Mg_3Cr_2Al_{18}$ or $CeCr_2Al_{20}$ structure,[13,14] in which V atoms are located at the 16$d$ position and Al atoms at the 96$g$, 48$f$, and 16$c$ positions, as depicted in Fig. 1. The A atom exists at the 8$a$ position with the $T_d$ site symmetry at the center of a large cage made of four 16$c$ Al and twelve 96$g$ Al atoms, which is a CN16 Frank-Kaspar polyhedron. Caplin and Nicholson found that $x$ = 0.3-0.7 for A = Al.[15] They also reported that the 8$a$ position was partially occupied by 0.6 Al and 0.1 Ga atoms in a Ga-doped sample. There are always vacancies at the 8$a$ position for small A atoms like Ga and Al, while no vacancy for large ones like Y, La, and Lu.[16]

Caplin and coworkers found unique local modes associated with small A atoms that are loosely bound to their neighbor and can move around inside the surrounding cage with low frequencies.[15,17] These local vibrations were characterized by Einstein modes with low Einstein temperatures: $T_E$ = 23 and 10 K for A = Al and Ga atoms, respectively. Cooper analyzed their resistivity data and found a characteristic temperature dependence caused by a scattering by the same Einstein mode of $T_E$ = 23 K.[18] Legg and Lanchester found an anomalously large lattice expansion in $A_xV_2Al_{20}$, which was ascribed to the volume dependence of local mode excitations.[19] On one hand, Claeson and Ivarsson reported superconductivity at $T_c$ = 1.6-1.7 K for compounds containing either Al or Ga atom inside the cage without giving any experimental data.[20] They mentioned that there was little evidence to show the contribution of the low-energy modes to the superconductivity.

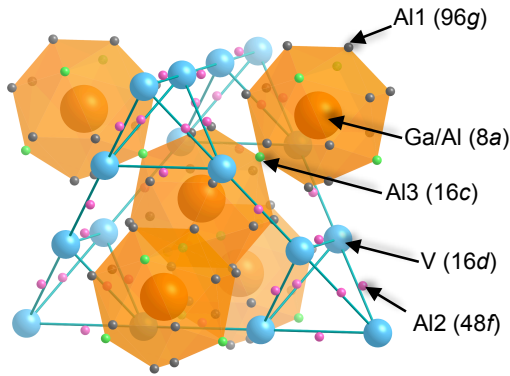

Fig. 1. (Color online) (a) Crystal structure of $A_xV_2Al_{20}$ in the $CeCr_2Al_{20}$ structure with space group $Fd\bar{3}m$ (origin choice 2). A rattling A (Ga or Al) atom at the $8a$ position is confined in a cage made of four $16c$ Al and twelve $96g$ Al atoms (CN16 Frank-Kaspar polyhedron).

Recently we have revisited the compounds in the light of the recent understanding of rattling and e-r interactions.[16] We synthesized a series of polycrystalline samples of $A_xV_2Al_{20}$ with $A_x$ = $Ga_{0.2}$, $Al_{0.3}$, Y, and La, the metallic radius of the A atoms being increased from Ga to La. A weak-coupling BCS superconductivity was observed at $T_c$ = 1.49, 1.66, and 0.69 K for $A_x$ = $Al_{0.3}$, $Ga_{0.2}$, and Y, respectively, but not above 0.4 K for $A_x$ = La. $T_c$ seems to increase with decreasing Einstein temperature, just as observed in β-pyrochlore oxides.[3] It has been shown that there is little but significant effect of low-energy modes or rattling on the electronic properties of the A = Al and Ga compounds. At nearly the same time, Safarik et al. also paid attention to this compound.[21] They studied $Al_{10.1}V$ ($Al_{0.2}V_2Al_{20}$) and gained evidence of a large anharmonicity for Al rattling. In the present study, we investigate the Ga content dependences of the lattice and electronic properties of $Ga_xV_2Al_{20}$, which have not been examined systematically but would provide us with an opportunity to gain better understanding of the relation between rattling and superconductivity in this class of cage compounds.

## 2. Experimental

Polycrystalline samples of $Ga_xV_2Al_{20}$ were prepared by a solid-state reaction from gallium, aluminum, and vanadium in nominal ratios of $x$ = 0.05-0.60. First, a composite was melted at a high temperature in an arc-melt furnace to obtain a uniform mixture. After sealing in an evacuated quartz ampoule, the mixture was annealed at 600-650°C for 80 h for the compound to form via a peritectic reaction.

Powder X-ray diffraction (XRD) measurements with a Cu-Kα radiation in RINT-2000 (Rigaku) indicate that almost monophasic samples were obtained for all $x$ values, as shown in Fig. 2(a). The small amounts of coexisting impurity phases are $Al_{45}V_7$ for a small $x$ and aluminum for a large $x$, which are expected to coexist with the main phase in the thermodynamic phase diagram as a result of slight overheating or a small deviation in the starting composition. However, it may not be harmful to investigate the bulk properties in the present study. A systematic variation in the Ga content is evidenced by gradual variations in the XRD profile. For example, the intensity ratio of the (113) reflection to the (222) reflection varies systematically with $x$, reflecting the difference in the atomic scattering factor between Al and Ga.

The lattice constant of the cubic unit cell determined at room temperature is listed in Table I and shown in Fig. 2(b) as a function of the Ga content. The actual lattice constant of $Ga_xV_2Al_{20}$ may be insensitive to $x$ because of the small relative change in composition and of the characteristic structure with a rigid cage. In fact, it is close to that of a pure compound of $Al_{0.3}V_2Al_{20}$ for $x$ < 0.20. However, it suddenly increases at $x$ = 0.20. The larger lattice constants for $x \geq 0.20$ may be associated with larger thermal expansions due to enhanced anharmonicity, as will be discussed later. The chemical composition was examined by inductively coupled plasma atomic-emission spectroscopy in a commercial equipment (JY138KH, HORIBA). A sample for the analysis was prepared by dissolving a pellet of approximately 5 mg into a dilute mixture of nitric acid and hydrochloric acid. The results confirmed that the intended compositions were retained in the products within an experimental resolution smaller than 1%, which corresponds to < 0.2 per formula unit (fu).

Magnetic susceptibility measurements between 2 and 300 K were employed in a Magnetic Property Measurement System (Quantum Design). Heat capacity was measured between 0.4 and 300 K by the relaxation method using a thin pellet of approximately 2 mg in a Physical Property Measurement System (Quantum Design). The data between 0.4 and 2 K were obtained using a $^3$He refrigerator.

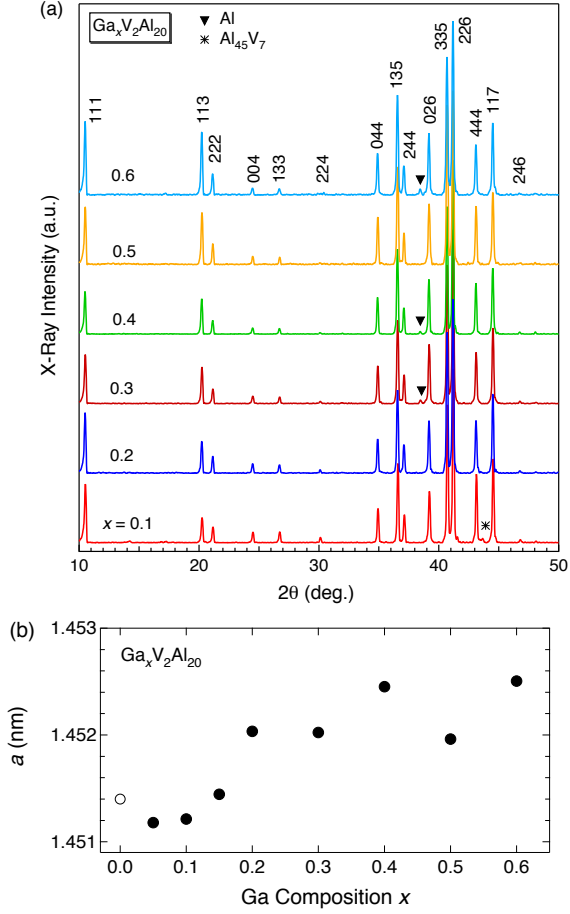

Fig. 2. (Color online) (a) X-ray diffraction profiles of polycrystalline samples of Ga$_x$V$_2$Al$_{20}$ with $x$ = 0.10-0.60. A Cu-K$\alpha$ radiation was employed. Each intensity is normalized to the peak height of the most intense 226 reflection. Indices of reflections on the basis of a cubic unit cell are given in the top profile for $x$ = 0.60. Small traces of impurity phases are observed: Al$_{45}$V$_7$ for a small $x$ and Al for a large $x$. (b) Ga content dependence of the lattice constant $a$. That of Al$_{0.3}$V$_2$Al$_{20}$ is plotted at $x$ = 0. The error bar is small within each mark.

## 3. Results
### 3.1. Superconductivity

Superconducting transitions in heat capacity and resistivity are shown for a typical composition of $x$ = 0.20 in Fig. 3. At zero magnetic field, a sharp drop in resistivity is observed at 1.76, 1.69, and 1.64 K at the onset, midpoint, and offset of the transition, respectively. Correspondingly, a second-order transition in heat capacity appears at 1.67, 1.65, and 1.62 K. Thus, zero resistivity occurs at around the midpoint of the jump in heat capacity, which is defined as a mean-field $T_c$. The superconductivity is completely suppressed above 0.4 K at a magnetic field of 0.3 T.

Generally, heat capacity consists of electronic ($C_e$) and lattice ($C_l$) contributions. Provided that $C_e$ is proportional to the Sommerfeld coefficient $\gamma$ in the normal state and that $C_l$ is independent of magnetic field, $C(0) - C(H) = C_e^s - C_e^n = C_e^s - \gamma T$, where $C(H)$ is the heat capacity at a magnetic field $H$ larger than $H_{c2}$ and $C_e^s$ and $C_e^n$ are the electronic heat capacities of superconducting and normal states. We fit $C(0) - C(1\,\text{T})$ divided by $T$ below $T_c/2$ to the form of $\exp(-\Delta/k_B T)$, taking into account the entropy balance, i.e.,

$$\int_0^{T_c} \frac{[C(0) - C(H)]}{T} dT = 0. \quad (1)$$

The results are satisfactory, as shown in Fig. 3(b), yielding $\Delta/k_B$ = 2.41 K or $2\Delta/k_B T_c$ = 2.92 and $\gamma$ = 35.0 mJ K$^{-2}$ mol$^{-1}$. The magnitude of the jump at $T_c$, $\Delta C/k_B T_c$, is 1.35. The same analyses were performed on all the samples, and the obtained parameters are summarized in Table I.

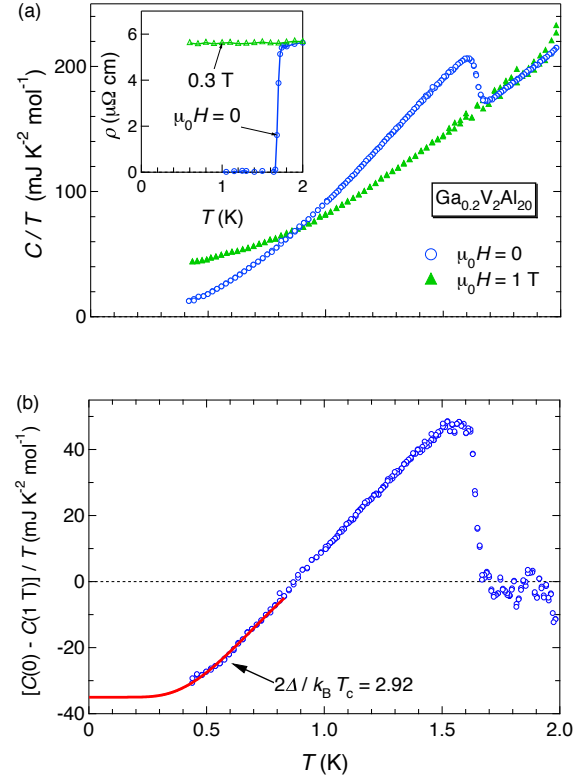

Fig. 3. (Color online) (a) Heat capacity divided by temperature $C/T$ and resistivity of Ga$_{0.2}$V$_2$Al$_{20}$ measured at zero field and magnetic fields of 1 and 0.3 T. (b) Electronic heat capacity obtained by subtracting the 1 T data from the zero-field data in (a), showing a superconducting transition with $T_c$ = 1.65 K and $\Delta C/\gamma T_c$ = 1.35 with $\gamma$ = 35 mJ K$^{-2}$ mol$^{-1}$. The curve below $T_c/2$ is a fit to $\exp(-\Delta/k_B T)$, yielding $2\Delta/k_B T_c$ = 2.92.

Figure 4 shows superconducting transitions observed in heat capacity. The jumps at $T_c$ remain sharp irrespective of the Ga content. The heat capacity of the $x = 0.05, 0.10,$ and $0.15$ samples are almost identical to each other and resembles that of $Al_{0.3}$ with a slightly higher $T_c$ by 0.04 K. In contrast, there are much larger phonon contributions above $T_c$ for samples with $x \geq 0.20$, indicating a substantial change in the phonon spectrum at $x \sim 0.20$. The $T_c$ is 1.65 K for the $Ga_{0.2}$ sample, 0.16 K higher than that of $Al_{0.3}$, and decreases gradually with further increasing $x$. The $x$ dependence of $T_c$ is shown in Fig. 5. Markedly, there is a distinct jump in $T_c$ by 8% between 0.15 and 0.20. Moreover, the $T_c$ is almost constant at $x < 0.20$, while it decreases linearly at $x \geq 0.20$.

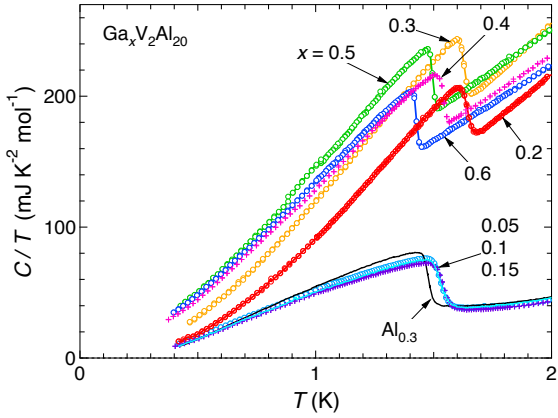

Fig. 4. (Color online) Heat capacity divided by temperature of $Al_{0.3}V_2Al_{20}$ and $Ga_xV_2Al_{20}$ with $x = 0.05, 0.10, 0.15, 0.20, 0.30, 0.40, 0.50,$ and $0.60$ measured at zero magnetic field.

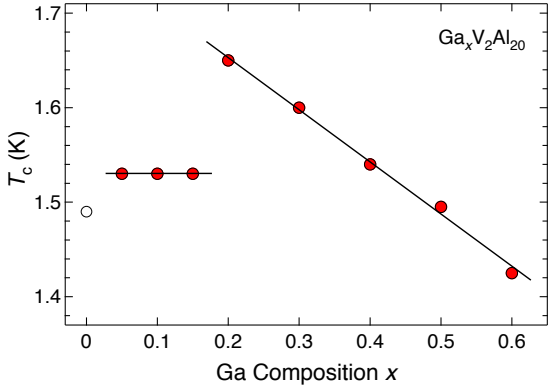

Fig. 5. (Color online) $T_c$ as a function of the Ga content. The point at $x = 0$ is from $Al_{0.3}V_2Al_{20}$. The lines are guides for the eyes.

Figure 6 shows a comparison of $C_e/T$ normalized by $\gamma$ for selected compositions. The curves nearly overlap with each other, showing a typical form for a weak-coupling BCS superconductivity. The magnitudes of the jump at $T_c$ are 1.35-1.41 (Table I), close to the theoretical value of 1.43. Moreover, $2\Delta/k_BT_c = 2.5$-$3.3$, slightly smaller than the theoretical value of 3.53. Thus, an $s$-wave superconductivity occurs, irrespective of the Ga content, as generally expected for a phonon-based superconductivity.

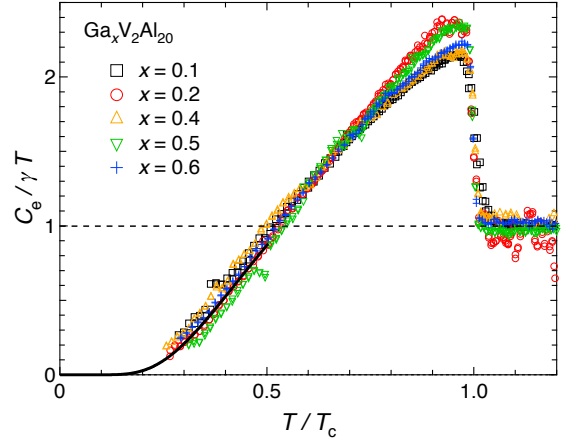

Fig. 6. (Color online) Electronic heat capacity divided by temperature and the Sommerfeld coefficient $\gamma$ as functions of normalized temperature $T/T_c$ for $Ga_xV_2Al_{20}$ with $x = 0.10, 0.20, 0.40, 0.50,$ and $0.60$. The values of $T_c$ and $\gamma$ are listed in Table I. The line is the fit shown in Fig. 3(b) for the $Ga_{0.2}$ data.

The upper critical field $H_{c2}$ is determined using heat capacity data measured under magnetic fields. We have noticed that, under magnetic fields, zero resistivity tends to occur at a higher temperature than the mean-field $T_c$ from the heat capacity. The origin of this is not clear but it may be the surface or interface superconductivity having a larger upper critical field than the bulk.[22] Figure 7(a) shows the heat capacity measured under various magnetic fields for $x = 0.10$ and 0.20, from which $H_{c2}$ is determined and plotted in an $H$-$T$ phase diagram in Fig. 7(b). The temperature dependence of $H_{c2}$ seems to follow a quadratic form as usually observed and can be reproduced by the empirical formula $H_{c2}(T) = H_{c2}(0)[1 - \alpha(T/T_c)^\beta]$: the parameters obtained are $[H_{c2}(0) / \text{mT}, \alpha, \beta] = [65, 1, 1.91], [109(1), 0.99(2), 1.7(3)],$ and $[137(7), 0.994(9), 1.8(2)]$ for $x = 0.10, 0.20,$ and $0.60$, respectively. These $H_{c2}$ values are much larger than the critical field of Al metal (9.9 mT) which is a type-I superconductor, suggesting that the present compounds are type-II superconductors. The Ginzburg-Landau coherence length $\xi$ is obtained from the equation $H_{c2}(0) = \phi_0/(2\pi\xi^2)$, where $\phi_0$ is the flux quantum: $\xi = 70, 55,$ and 50 nm, respectively. Note that the $H_{c2}$ increases with increasing $x$, which may be attributed to a higher disorder caused by the Ga substitution.

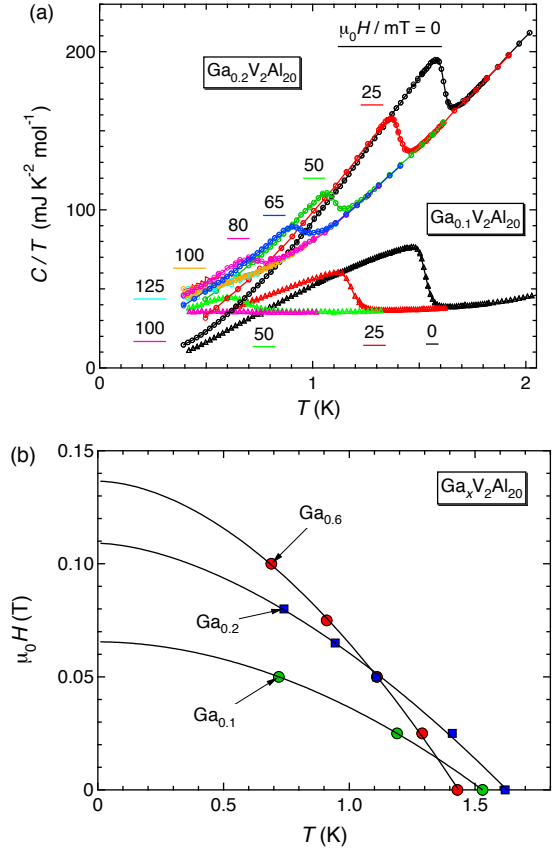

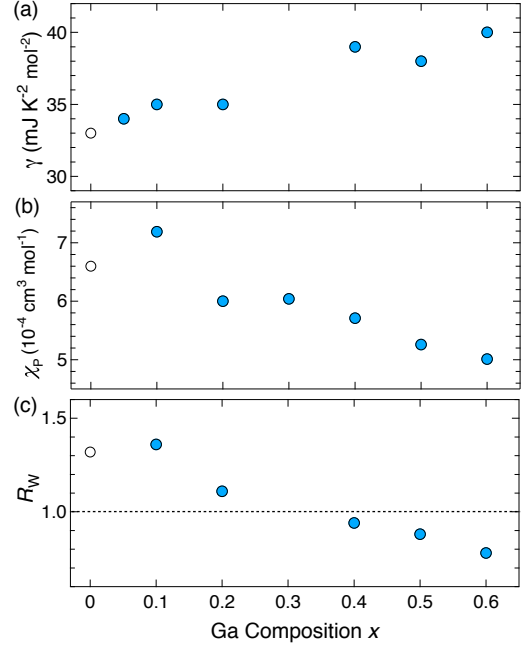

Fig. 8. (Color online) Ga content dependences of (a) Sommerfeld coefficient $\gamma$, (b) Pauli paramagnetic susceptibility $\chi_P$, and the Wilson ratio $R_W$. Plotted at $x = 0$ are those of $Al_{0.3}V_2Al_{20}$.

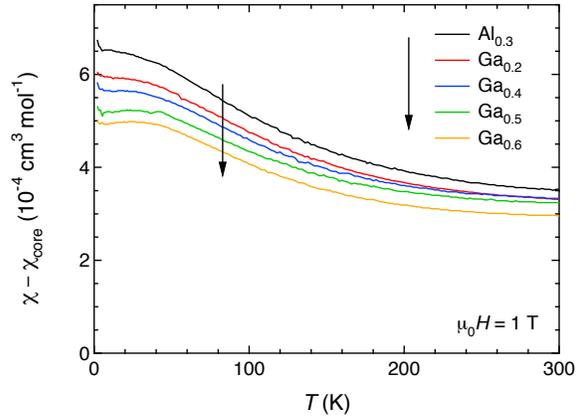

Fig. 7. (Color online) (a) Heat capacities of the $Ga_{0.1}$ and $Ga_{0.2}$ samples measured under various magnetic fields. (b) Magnetic field-temperature phase diagram. The upper critical field determined by heat capacity measurements is plotted for $x = 0.10$, 0.20, and 0.60. The curve on each dataset is a fit to the form of $H_{c2}(T) = H_{c2}(0)(1 - \alpha(T_c/T)^\beta)$, which yields $H_{c2}(0) = 65$, 109, and 137 mT for $x = 0.10$, 0.20, and 0.60, respectively.

### 3.2. Normal-state properties

The Ga content dependence of the Sommerfeld coefficient $\gamma$ is shown in Fig. 8(a). $\gamma$ increases gradually with increasing $x$, which does not reflect the corresponding variation in $T_c$ shown in Fig. 5; $\gamma$ gives a measure of the renormalized density of states (DOS) at the Fermi level as well as of the electron-phonon (e-ph) coupling strength and should scale with $T_c$ to some extent in a simple BCS picture. This seemingly contradicting behavior will be discussed later.

Another experimental measure of the electronic DOS is magnetic susceptibility, which is proportional to DOS in the free-electron theory and can be enhanced in the presence of electron correlations in the Fermi liquid theory. Figure 9 shows the magnetic susceptibilities of selected samples, all of which increase gradually upon cooling and tend to saturate below ~20 K. This characteristic temperature dependence may be due to spin fluctuations arising from electron correlations or due to the band smearing effect which occurs when the Fermi level is located near a peak in DOS. The Pauli paramagnetic susceptibility $\chi_P$ is decided as the value at 10 K after the subtraction of contributions from core diamagnetism (approximately $-6.6 \times 10^{-5}$ cm$^3$ mol$^{-1}$ for all compositions). With increasing $x$, $\chi_P$ decreases systematically [Fig. 8(b)], which probably means a reduction in DOS. Then, the Wilson ratio $R_W = (\pi^{2/3})(k_B/\mu_B)^2(\chi/\gamma)$ decreases with increasing $x$ [Fig.

Fig. 9. (Color online) Magnetic susceptibility measured at $\mu_0H = 1$ T after subtraction of the core diamagnetism ($\chi_{core} \sim -6.6 \times 10^{-5}$ cm$^3$ mol$^{-1}$ for all contents).

8(c)]. $R_W$ = 1.32 for Al$_{0.3}$ suggests a moderately strong electron correlation, while $R_W < 1$ for $x \geq 0.4$, e.g., $R_W$ = 0.78 for Ga$_{0.6}$, means that electron-phonon interactions are significantly enhanced with increasing $x$; electron-phonon interactions do not increase $\chi_P$ but $\gamma$. The origin of the enhancement of electron-phonon interactions may be related to rattling. Actually, very small $R_W$s have been reported for β-pyrochlore oxides AOs$_2$O$_6$; $R_W$ = 0.48 for A = Cs decreases to 0.14 for K with more intense rattling.[7]

### 3.3. Band structure calculations

Electronic band structures have been calculated in order to get insight into the effects of band filling for A$_x$V$_2$Al$_{20}$. Calculations were carried out by a full-potential APW method with the local density approximation for the exchange correlation potential. Two extreme cases with 0 and 100% Al atoms at the 8$a$ site are examined, assuming the same structural parameters as those of Al$_{0.3}$V$_2$Al$_{20}$.[13] Reflecting the large unit cell and the high symmetry of the crystal structure, the band structures are rather complex and give DOS with many sharp spikes, as shown in Fig. 10. There are 140 and 146 electrons per primitive cell for V$_2$Al$_{20}$ and AlV$_2$Al$_{20}$, respectively. In order to clarify the effect of electron filling between them, the $E_F$ of AlV$_2$Al$_{20}$ is not set to the origin of energy, but shifted by 0.302 eV above the $E_F$ of V$_2$Al$_{20}$, so that filling by 140 electrons corresponds to $E = 0$ in both compounds. In each case, the Fermi level is located near a sharp peak. The DOSs at $E_F$ are 14.996 and 12.306 states eV$^{-1}$ fu$^{-1}$, which correspond to $\gamma_{band}$ = 35.3 and 29.0 mJ K$^{-2}$ mol$^{-1}$, respectively.

An actual electron filling for A$_x$V$_2$Al$_{20}$ should lie between AV$_2$Al$_{20}$ and V$_2$Al$_{20}$. If extra electrons from A$_{0.3}$ atoms were simply added to V$_2$Al$_{20}$ in a rigid band picture, $E_F$ would be raised by 0.087 eV and lie near a steep valley in the DOS of V$_2$Al$_{20}$. Then, an expected $\gamma_{band}$ would be 8.5 mJ K$^{-2}$ mol$^{-1}$, which is too small compared with the experimental value of 33 mJ K$^{-2}$ mol$^{-1}$ for Al$_{0.3}$V$_2$Al$_{20}$; the mass enhancement factor $\gamma_{exp}/\gamma_{band}$ should be smaller than 1.5 for a weak-coupling superconductor. Thus, such a rigid band picture is not rationalized, although the reduction in DOS is likely to occur to some extent for electronic stabilization. The failure of a rigid band picture is inferred from the comparison of the two DOS profiles in Fig. 10: they are considerably different in spite of the small change in composition and the same structure parameters used in the calculations. This is mainly because there is a significant contribution of the 8$a$ Al atom to the electronic states near the Fermi level, as shown by the partial DOS at the bottom of Fig. 10(a); even a partial occupancy of the 8$a$ site may change the DOS profile markedly. Thus, unfortunately, it is difficult to deduce the mass enhancement in the present compounds by band structure calculations. Nevertheless, note that the A atom possesses a strong nonionic character in A$_x$V$_2$Al$_{20}$, which is in contrast to the case of AOs$_2$O$_6$ with the A atom having a completely ionic nature.

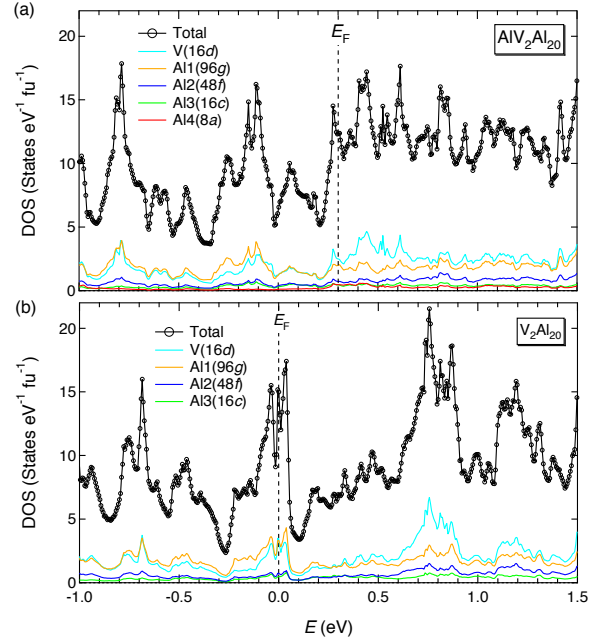

Fig. 10. (Color online) Calculated DOSs for (a) AlV$_2$Al$_{20}$ and (b) V$_2$Al$_{20}$ with the 8$a$ site completely occupied and vacant, respectively. Total and partial DOSs from each atom are shown. The $E_F$ of AlV$_2$Al$_{20}$ is set to 0.302 eV, so that the same electron filling as V$_2$Al$_{20}$ takes $E = 0$.

### 3.4. Rattling in heat capacity

Low-energy rattling modes manifest themselves at low temperatures in heat capacity. Clearly, there are large enhancements below 30 K in Ga$_x$ compounds than in LaV$_2$Al$_{20}$, as shown in Fig. 11. The $C/T$ of the La compound is proportional to $T^2$ below ~10 K, indicating that the lattice heat capacity at low temperatures is well described by the Debye model at a reasonable Debye temperature of $T_D$ = 430 K; the vibration of La atoms takes a large energy and may be incorporated into normal lattice vibrations. A common contribution to heat capacity of the V$_2$Al$_{20}$ framework can be estimated from the $C/T$ of LaV$_2$Al$_{20}$ after the subtraction of $\gamma$ and the multiplication by a factor of 22/23. Note that all the data of the Ga$_x$ samples above 100 K are nearly equal to each other and well reproduced by the framework contribution estimated from the La compound. This implies that most phonon modes, i.e., 129 optical modes for V$_2$Al$_{20}$, are almost identical as naturally expected. In the case of Al$_{10}$V, Caplin and Nicholson found an acoustic mode with $T_D$ = 420 K.[15]

The contribution of A-atom vibrations $C(A)/T$ after the subtraction of the cage contribution exhibits a broad peak at 7-10 K, as shown in Fig. 12. The $C(A)/T$ of the $Ga_{0.1}$ sample resembles that of the $Al_{0.3}$ sample[16] and is well reproduced by assuming a single Einstein mode with a number of oscillators $\delta = 0.30(1)$ per fu and an Einstein temperature $T_E = 24.2(1)$ K; $\delta = 0.280(3)$ and $T_E = 23.7(2)$ K for $Al_{0.3}$. This suggests the lack of contribution of the rattling of Ga atoms; Ga atoms must have replaced Al atoms at the framework by pushing 0.3 Al atoms inside the cage. On the other hand, the peaks in $C(A)/T$ for $x \geq 0.2$ are expanded to lower temperatures than that of $Ga_{0.1}$. The $Ga_{0.2}$ data cannot be reproduced by assuming a single Einstein mode but by assuming two Einstein modes, i.e. $(\delta, T_E) = (0.050(2), 8.1(1)$ K$)$ and $(0.250(3), 23.4(3)$ K$)$. The former is presumably due to a smaller Ga atom contained in the cage and the latter due to an Al atom. This means that 0.05 Ga atoms are present concomitantly with 0.25 Al atoms in the cage, and the rest of the Ga atoms, i.e. 0.15, should occupy the cage position. Note that the large enhancement in the low-temperature heat capacity of $Ga_{0.2}$ compared with that of $Ga_{0.1}$ is mostly due to this small amount of Ga atoms in the cage.

The numbers of oscillators $\delta(A)$ and the Einstein temperatures $T_E(A)$ are summarized in Table I and plotted as functions of the Ga content in Fig. 13. Samples of $x < 0.20$ contain only Al atoms of approximately 0.3 per fu in the cage. $\delta(Al)$ remains almost the same as $x$ increases over 0.20. On the other hand, Ga atoms are suddenly introduced into the cage above $x = 0.20$, always keeping $\delta(Ga)$ near 0.05, indicating that most Ga atoms replace Al atoms forming the cage. There may be some reason for the "magic" numbers $\delta(Al) = 0.25$-$0.35$ and $\delta(Ga) \sim 0.05$ or $[\delta(Al) + \delta(Ga)] \sim 0.3$. One possibility is that the electronic structure becomes stable around this electron filling. However, it is difficult to confirm this in our band structure calculations, as mentioned before. As will be described later, the effects of Ga vibrations on the electronic properties of $Ga_xV_2Al_{20}$ become obvious by examining the changes in various parameters across the boundary at $x = 0.02$.

One notable feature in Fig. 12 is that Einstein modes cannot reproduce the data below 2 K for $x \geq 0.20$, which is in contrast to the case of $x = 0.10$ with an almost perfect fitting. This strongly suggests that Ga rattling seriously affects the acoustic phonon modes of the cage. As observed in Fig. 11(c), the $C/T$ data for $x \geq 0.20$ are almost proportional to $T^2$ at low temperatures below ~1 K with coefficients much larger than that of $Ga_{0.1}$ or $LaV_2Al_{20}$. The Debye temperatures deduced from the coefficients are ~90 K (Table 1), which are much smaller than $T_D = 430$ K for $Al_{0.3}$ ($Ga_{0.1}$) and $LaV_2Al_{20}$. Thus, the Ga vibration gives rise to an unusually large softening of the cage framework, in spite of the fact that the number of Ga atoms in the cage is merely 0.05.

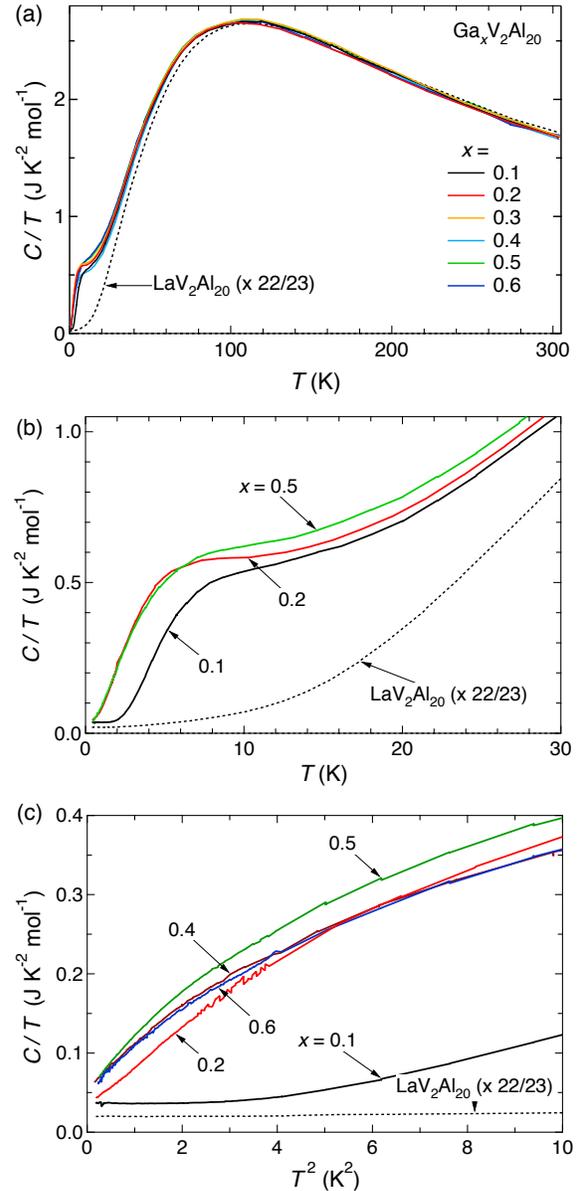

Fig. 11. (Color online) Heat capacity divided by temperature in a wide temperature range (a), below 30 K (b), and as a function of $T^2$ at low temperatures (c). That of $LaV_2Al_{20}$ after the subtraction of $\gamma$ and the multiplication by a factor of 22/23 is also shown as reference for the contribution of the cage framework.

It is interesting to note in Fig. 2(b) that there is a jump in the lattice constant at $x = 0.15$-$0.20$. This is obviously due to the introduction of Ga atoms into the cage. Since the lattice constant has been determined at room temperature, it is likely that thermal expansion is enhanced by the extremely low-energy vibrations of Ga atoms; the cage becomes softer and the anharmonicity is enhanced. In fact, a large lattice expansion has been observed for Al$_x$ compounds and ascribed to the anharmonicity of acoustic phonons induced by Al rattling.[19] Since the Ga rattling has a much lower energy and must affect the acoustic phonons more seriously, a larger lattice expansion is expected with the Ga rattling.

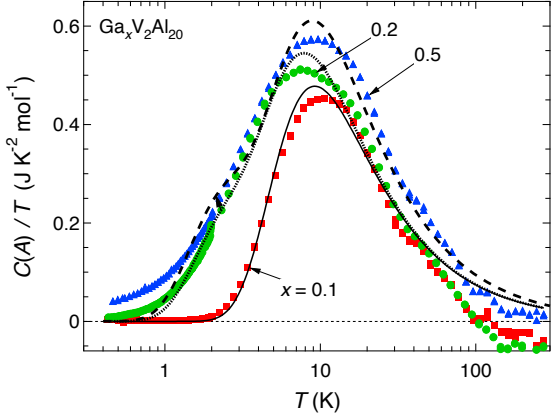

Fig. 12. (Color online) Heat capacity of A atom vibrations inside the cage, $C$(A), after subtraction of the cage contribution estimated from the data of LaV$_2$Al$_{20}$. The curve for $x = 0.10$ represents a fit assuming a single Einstein mode, and those for $x = 0.20$ and $0.50$ are fits assuming two Einstein modes. The results of the fitting are summarized in Fig. 13 and Table I.

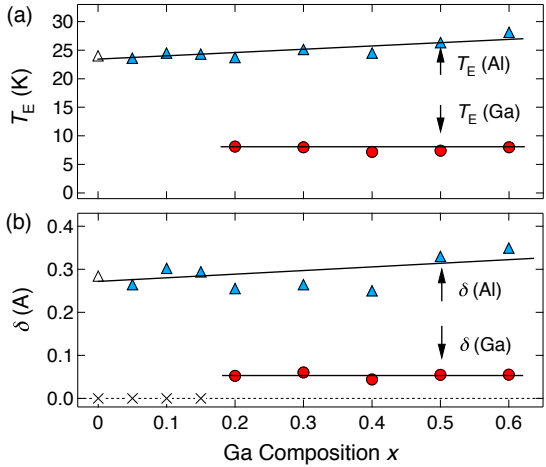

Fig. 13. (Color online) Ga content dependences of Einstein temperatures (a) and the numbers of oscillators per fu for Al and Ga (b). $\delta$(Ga) is assumed to be zero for $x < 0.20$. The lines are guides for the eyes.

### 3.5. Electrical resistivity

A possible interaction between the A-atom vibrations and conduction electrons should manifest itself in the temperature dependence of electrical resistivity at low temperatures, because such low-energy modes can give rise to a large scattering of carriers even at low temperatures where few normal phonons are generated. It has been pointed out that a local mode can be effective at the lowest temperatures, because it should scatter electrons almost isotropically; bulk phonons can do this at only small angles.[15] The resistivity curves of the Al$_{0.3}$ and Ga$_x$ compounds are shown in Fig. 14, together with those of LaV$_2$Al$_{20}$[16] and β-pyrochlore oxides.[7] The $\rho$s of polycrystalline samples of Al$_{0.3}$ and Ga$_x$ are 60-150 μΩ cm at 300 K and 5-15 μΩ cm just above $T_c$, with RRR ~ 10.

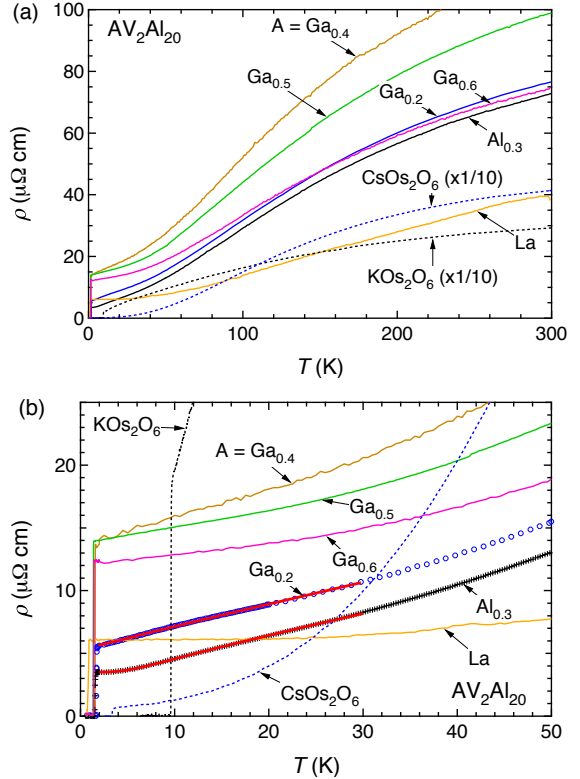

Fig. 14. (Color online) Temperature dependences of resistivity for polycrystalline samples of A = Al$_{0.3}$, Ga$_x$, and La. Those of single crystals of two β-pyrochlore oxides, i.e., CsOs$_2$O$_6$ and KOs$_2$O$_6$, are also shown for comparison. Solid lines below 30 K on the datasets of Al$_{0.3}$ and Ga$_{0.2}$ in (b) show fits to the formula given by Cooper,[18] assuming single and double Einstein modes, respectively.

At low temperatures, the La compound and CsOs$_2$O$_6$ show $T^3$ and $T^2$ dependences, respectively. In contrast, the Al$_{0.3}$ and Ga$_x$ compounds exhibit stronger $T$ dependences, i.e., an almost $T$-linear behavior below ~20 K for Ga$_x$ with a large $x$. Cooper analyzed the $\rho$ data of Al$_x$V$_2$Al$_{20}$ by Caplin et al. by assuming a scattering by an Einstein mode as

$$\rho = \frac{KN}{T(e^{\Theta/T} - 1)(1 - e^{-\Theta/T})}, \qquad (2)$$

where $N$ is the number of oscillators and $K$ is a constant.[18] He obtained a good fit with $T_E = 22$ K.

Using eq. (2), we analyze our data of $Al_{0.3}$ and obtain a good fit with $T_E$ = 23.6(3) K, as shown in Fig. 14(b), which is in perfect agreement with $T_E$ = 23.7(2) K from heat capacity. On the other hand, the $\rho$ data of $Ga_x$ cannot be fitted with eq. (2) with a single Einstein mode. Taking into account of the heat capacity data, we fit the $Ga_{0.2}$ data to the form of $\rho = \rho(Ga) + \rho(Al)$ with each contribution taking the same form as eq. (2). Assuming 0.05 Ga and 0.25 Al atoms in the cage, we obtain $T_E$ = 5.1 and 13.2 K, respectively. The fitting is satisfactory though the $T_E$s are smaller than those from the heat capacity. More reliable evaluations for such low-energy phonons would become possible with appropriate $\rho$ data down to much lower temperatures than $T_E$s. Electron-rattler interactions in $A_xV_2Al_{20}$ are weak but in fact present, serving as scatterers even at such low temperatures.

## 4. Discussion

The main focus of the present study is the electron-rattler interaction and its effect on the superconductivity in $A_xV_2Al_{20}$. In a previous study, it was suggested that there is little evidence of the relationship between rattling modes and superconductivity.[20] It is clear that rattling does not play a major role in the mechanism of superconductivity, which is in contrast to the case of β-pyrochlore oxides.[3] On the basis of the present results on $Ga_xV_2Al_{20}$, however, we point out that there is a certain correlation between them, although small. Here, we discuss superconductivity and the characteristics of rattling in reference to other related cage compounds.

### 4.1. Superconductivity in $Ga_xV_2Al_{20}$

First, note the typical weak-coupling superconductivity of pure aluminum metal. It occurs at $T_c$ = 1.16 K with $\gamma$ = 1.35 mJ K$^{-2}$ mol$^{-1}$ and $T_D$ = 428 K.[23] These parameters are quite similar to those of the present compounds. In the simplest form according to the BCS theory, $T_c$ is given as

$$T_c = \langle\omega\rangle\exp(-1/\lambda), \quad (3)$$

with $\lambda = N(0)V$, where $N(0)$ is the DOS at the Fermi energy and $V$ is the pairing potential arising from the e-ph interaction using a typical phonon energy $\langle\omega\rangle$. Provided that $\langle\omega\rangle = 1.13T_D$, $\lambda$ is obtained as 0.165 for Al.

Let us consider two typical compounds: $Al_{0.3}V_2Al_{20}$ and $Ga_{0.2}V_2Al_{20}$. There is a small but distinct difference in $T_c$ between them, as shown in Fig. 5, which must be related to Ga rattling. Nevertheless, it is reasonable to assume that these superconductivities are primarily induced by normal phonons, as in pure Al, because they are weak-coupling superconductors. As shown in the heat capacity in Fig. 11, high-energy phonons are almost identical between the two compounds, with the typical energy being $T_D$ = 420 K.[15] Then, $\lambda$ is estimated as 0.173 and 0.176 for $Al_{0.3}$ and $Ga_{0.2}$, respectively, from eq. (3). Thus, the rise of $T_c$ by 8% at $x \sim 0.20$ is due to this tiny increase in the e-ph interaction. A corresponding change in $\gamma$ would be too small to observe (Fig. 8).

Although low-energy modes do not have major roles, they must serve to enhance the e-ph interaction. $\lambda$ is given by

$$\lambda = 2\int_0^\infty d\omega \frac{\alpha^2(\omega)F(\omega)}{\omega}, \quad (4)$$

where $\alpha^2(\omega)$ is the coupling function and $F(\omega)$ is the phonon DOS. Thus, a low-energy phonon should enhance $\lambda$ effectively: the low-energy part of $\alpha^2(\omega)F(\omega)$ becomes dominant owing to the denominator $\omega$. This is the reason why many strong-coupling superconductors possess low-energy phonons, as in the A-15[24] and Chevrel phase compounds.[25]

The phonon DOS for the present compounds at a low energy is schematically illustrated in Fig. 15. The $Ga_{0.1}$ compound, the same as $Al_{0.3}$, has a peak at $E/k_B$ = 23 K from the Al rattling mode above a small background from a Debye phonon with $T_D$ = 420 K. In contrast, an additional peak exists at 8 K in $Ga_{0.2}$ from Ga rattling. Moreover, the Debye tail rises up rapidly because of the small $T_D$ = 90 K. Consequently, the enlarged phonon DOS at a low energy results in the observed enhancement in $\lambda$ for $Ga_x$ with $x \geq 0.20$. However, since the relative weight of this low-energy part is rather small in the overall phonon DOS, the increases in $\lambda$ and $T_c$ remain minimal and the superconductivity remains in the weak-coupling regime.

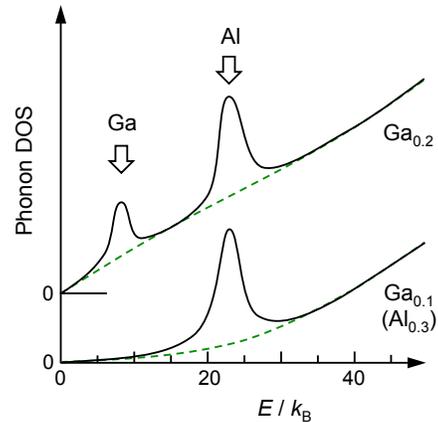

Fig. 15. (Color online) Schematic representation of the phonon DOSs for the $Ga_{0.1}$ ($Al_{0.3}$) and $Ga_{0.2}$ samples. Contributions of acoustic modes are shown by broken

curves.

There is another feature in the $x$ dependence of $T_c$ in Fig. 5: $T_c$ decreases linearly with increasing $x$ above 0.20, in spite of the gradual increase in $\gamma$ or the enhancement of the e-ph interaction shown in Fig. 8. This is probably due to the randomness caused by the substitution of Ga for Al at the cage position. It is known for pure Al metal that nonmagnetic impurities can reduce $T_c$ linearly; $T_c$ becomes lower by 0.05 K at 1 at.%.[26] The change in $T_c$ in Fig. 5 is $\Delta T_c$ = -0.22 K from $x$ = 0.20 to 0.60, which means that $\Delta T_c$ = -0.11 K/at.% Ga, assuming that all the nominal Ga atoms out of the 0.05 atoms inside the cage occupy the Al positions. By extrapolating the line above $x$ = 0.20 to $x$ = 0.05 in Fig. 5, one would expect the intrinsic $T_c$ in the absence of disorder to be 1.74 K for an ideal compound of $(Ga_{0.05}Al_{0.25})V_2Al_{20}$ with all Ga atoms only inside the cage.

*4.2. Rattling in $A_xV_2Al_{20}$*

Caplin and Nicholson identified the low-energy mode of the guest Al4 atom as a unique local mode with a small energy spread of ~2 K.[15,17] This is because the energy is too low to hybridize with bulk phonon modes. They suggest that there are a local maximum at the cage center and four local minima 55 pm apart in the [111] directions in the tetrahedral symmetry.[15] Since the potential between off-center minima is much smaller than that of the on-center, the guest atom can move around most of the spherical inner surface of the cage as a "rotator". The lowest energies of this rotator are 23 and 8 K for Al and Ga atoms, respectively. On the other hand, Kontio and Stevens derived analytical expressions for an anharmonic one-particle-potential model and applied this to the Al4 atom in $Al_{0.84}V_2Al_{20}$.[27] They found an anharmonic thermal vibration of the on-center nature for the Al4 atom. More recently, Safarik *et al.* have also reported an on-center rattling for the Al4 atom in $Al_{0.2}V_2Al_{20}$ on the basis of their analysis of the temperature dependence of the atomic displacement parameter.[21] Thus, the issue of off-center or on-center vibrations still remains controversial in $A_xV_2Al_{20}$. An off-center rattling mode has been found in the clathrate $Eu_8Ga_{16}Ge_{30}$ with $T_E$ = 30 K, while an on-center mode for larger Ba atoms in $Ba_8Ga_{16}Ge_{30}$.[28] All the other compounds in the families of skutterudites and β-pyrochore oxides show on-center rattling.[3,29]

One marked feature of the present compounds is the partial occupancy of rattling atoms in the cage; the other cage compounds assume full or almost full occupancy. However, there is some controversy on the occupation of the Al4 atom. Previous XRD studies showed the occupancy factor $g$ to be $g$ = 0.1, 0.5, 0.84 and 0.90,[13,14,21,27] while heat capacity studies always found ~30% rattling atoms.[15,16,21] There may be experimental difficulties in determining the occupancy of the rattling site, because in general the occupancy and the atomic displacement parameter have a strong correlation. This may be more serious for a vibrating atom with a large anharmonicity, which makes it difficult to determine the occupancy correctly. It is also related to the issue of on-center or off-center rattling. There is an alternative possibility that the cage contains extra "fixed" atoms in addition to the 30% rattling atoms. However, this may not be the case, because the structural refinements did not detect extra electron density at off-center positions near the 8$a$ site.[13,14,27]

We think that the estimation on the basis of heat capacity is more reliable. According to our heat capacity analysis, the number of oscillators is always ~0.3. Plausibly, this magic number results from a structural constraint: the four neighboring cages around one cage occupied by a rattling Ga/Al atom tend to be vacant; a full occupation can be attained only for larger A atoms without rattling. Since this concentration is lower than the percolation limit for the diamond lattice, no cooperative phenomena are expected for the rattlers. Moreover, the density of rattling atoms among all atoms is as low as 1.5%.

*4.3. Rattling energy*

We consider the characteristic energies of rattling, i.e., 23 K for Al and 8 K for Ga atoms. The lower energy for a heavier Ga atom can be interpreted as a normal harmonic oscillation (the atomic masses of Al and Ga are 26.98 and 69.72, respectively). As pointed out in previous studies, however, a rattler should be treated as a particle in a box, where a potential takes a flat bottom near the center with a negligible spring constant and a steep wall far apart when the rattler hits cage-forming atoms.[30] Safarik *et al.* proposed a sixth-order potential for the Al4 atom in $VAl_{10.1}$.[21] The rattling energy has been sorted out in terms of the guest free space (gfs) in skutterudites and β-pyrochlores: the larger the gfs, the smaller the rattling energy.[31-33] gfs is defined as the distance from the center to the inner surface of the cage and is approximated by $d_{cage}$: $d_{cage}$ is a distance from the center A atom to the nearest-neighbor cage-forming atoms after the subtraction of the radius of the cage atom or ion.

The ratio $r_A/d_{cage}$ provides us with a key parameter to discuss the rattling energy from the structural viewpoint, where $r_A$ is the ionic or metallic radius (atomic radius in a metal) of the A atom, depending on the electronic state of the A atom; $r_A/d_{cage}$ ~ 1 for a normal crystal and can be smaller for a rattling A atom/ion. In the present compounds, it may be appropriate to assume metallic radii for Al and Ga atoms at the 8a site, because there are considerable electronic contributions of them to states near the Fermi level, as shown in the DOS in Fig. 10. The metallic radii are 135 and 143 pm for Ga and Al atoms, respectively. The bond length from the center to the nearest-neighbor Al atoms (four Al3) is 314 pm from the literature.[13] Then, $d_{cage}$ is 171 pm. Taking this $d_{cage}$ independent of the kind of rattling atom, $r_A/d_{cage} = 0.79$ and 0.84 for Ga and Al rattlers, respectively. Hence, the smaller the $r_A/d_{cage}$, the lower the rattling energy. For comparison, the $r_A/d_{cage}$ becomes 0.99 for A = Lu ($r_A$ = 174 pm) and 1.07 for La (187 pm). A large Lu atom fits the cage well and a larger La atom tends to expand the cage. In fact, the lattice constant at room temperature remains nearly constant or slightly decreases with increasing $r_A$ from A = $Al_{0.3}$ (1.45157 nm) to Lu (1.45084 nm) and becomes as large as 1.4622 nm for La, 0.73% larger than that of $Al_{0.3}$.

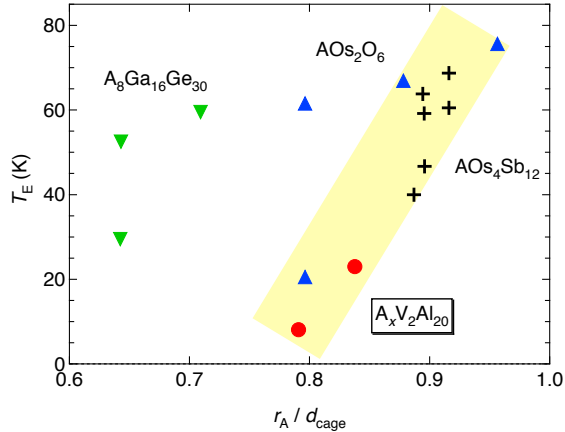

Fig. 16. (Color online) Einstein temperature vs $r_A/d_{cage}$ for $A_xV_2Al_{20}$ (A = Ga and Al), β-pyrochlore oxides $AOs_2O_6$ (A = Cs, Rb, and K), filled skutterudites $AOs_4Ab_{12}$ (A = La, Ce, Pr, Nd, and Sm), and clathrates $A_8Ga_{16}Ge_{30}$ (A = Ba, Sr, and Eu).

The relations between $T_E$ and $r_A/d_{cage}$ for various cage compounds are compared in Fig. 16. In the case of β-pyrochlores, the ionic radii are assumed for A = K, Rb, and Cs (they have no electronic contribution at $E_F$ and are completely ionic), and the $r_A/d_{cage}$ parameters are 0.80, 0.88, and 0.98, respectively.[31] Note that the value of $KOs_2O_6$ with $T_E$ = 22 K is close to that of $Ga_xV_2Al_{20}$. Skutterudite compounds with relatively large cages are $AOs_4Sb_{12}$ with A = La, Ce, Pr, Nd, and Sm. They take small $r_A/d_{cage}$ values at 0.61-0.67 for the ionic radii of the A atoms[32,33] and more reasonable values of 0.89-0.92 for metallic radii. The clathrate compounds $A_8Ga_{16}Ge_{30}$ with A = Ba, Sr, and Eu take very small values of 0.64-0.71 for the ionic radii, which implies that the parameter is not appropriate for these compounds with low-symmetry cages; possibly, off-center freezing takes place in these clathrate compounds, so that the actual $r_A/d_{cage}$ is larger.[28] It would be interesting to test how $T_E$ approaches zero for a smaller rattler, i.e., smaller than the Ga atom in $A_xV_2Al_{20}$ or the K ion in $AOs_2O_6$; unfortunately, there are no such small atoms/ions in nature that can be introduced in cages.

### 4.4. Softening of the cage

We point out the softening of the cage due to Ga rattling in $A_xV_2Al_{20}$. When only 30% Al atoms are rattling in the cage for $x < 0.20$, the cage is hard with $T_D$ = 420 K as in $LaV_2Al_{20}$.[15,16] In sharp contrast, the cage becomes much softer with $T_D$ = 90 K when 5% Ga atoms are added for $x \geq 0.20$. The lattice constant at room temperature increases at $x \geq 0.20$, suggesting a larger thermal expansion coming from enhanced anharmonicity.[19] The effect of the Ga substitution for the cage-forming Al must be minor, because the Ga substitution already occurs at $x < 0.20$. The sudden change in $T_D$ at $x = 0.20$ strongly suggests that Ga rattling causes this marked softening. It remains a question whether the softening is simply due to the low energy of Ga rattling or whether the rattling of Ga atoms is substantially different from that of Al, strongly coupled to the cage.

Safarik et al. found for $Al_xV_2Al_{20}$ that the Al3 atom at the 16c site, which is located just between adjacent Al4 atoms at the 8a site (Fig. 1), has a relatively large atomic displacement parameter and anharmonicity.[21] It is suggested that the Al3 atom vibrates anharmonically only when the neighboring 8a sites are occupied by a rattling Al atom. Note that the 16c site is closest to the 8a site and can have two neighbors at the 8a site, although the possibility to have two rattlers at the same time is quite small. The present results on $Ga_xV_2Al_{20}$ indicate that the softening of the cage due to the anharmonic vibration of the 16c Al atoms is still small for Al rattling and becomes enormously large for Ga rattling. This fact suggests that Ga rattling is quite different from Al rattling. A similar softening of the cage oxygen modes has been observed by Raman scattering experiments in $KOs_2O_6$, but not in the other β-pyrochlore oxides.[34] Rattling originally occurs with a weak coupling to the cage, but seems to shake the cage atoms seriously by hitting the potential wall when the energy becomes lower and the atomic excursion becomes larger.

### 4.5. Electron-rattler interactions

The electron-rattler interaction is apparently weak but present in the present compounds, as evidenced by the almost $T$-linear resistivity for higher Ga contents. In β-pyrochlore oxides, a large e-r coupling has been established, for example, by a large scattering of carriers showing $T^2$ resistivity with the coefficient enormously increasing with decreasing $T_E$ from Cs to KOs$_2$O$_6$.[3] Comparing Ga$_x$V$_2$Al$_{20}$ of $x \geq 0.20$ with KOs$_2$O$_6$, the $T_E$ of Ga rattling is lower than that of K rattling, suggesting more intense rattling vibrations. The question is which factors dominantly determine the magnitude of the e-r interaction. One of them may be the density of rattlers: only 0.2% per atom for Ga in A$_x$V$_2$Al$_{20}$ and 14% per atom for K in KOs$_2$O$_6$. However, we assume two more important factors: one is the ionicity of rattlers and the other the screening effect. In AOs$_2$O$_6$, the A atom is completely ionized with its electronic state located from $E_F$.[35,36] Moreover, conduction electrons come from Os 5$d$ states hybridized with O 2$p$ states, and thus, tend to be confined in the cage framework with little distribution near the center of the cage. This allows the A$^+$ ion inside the cage to strongly couple with conduction electrons around the cage through unscreened Coulomb interactions, particularly when it moves with a large excursion in a highly anharmonic potential. In contrast, the Ga or Al atom in the cage of A$_x$V$_2$Al$_{20}$ takes a nonionic character with their significant contributions near $E_F$ (Fig. 10). In addition, conduction electrons mostly from spread Al 2$s$/2$p$ states are widely distributed over the crystal; the screening effect should work effectively. Therefore, Coulomb interactions cannot be large in A$_x$V$_2$Al$_{20}$. A similar situation may be realized in skutterudite compounds, where e-r interactions are also small.

**5. Conclusions**

The rattling and superconductivity of a cage compound Ga$_x$V$_2$Al$_{20}$ are studied in a series of polycrystalline samples with $x$ = 0.05-0.60 and Al$_{0.3}$V$_2$Al$_{20}$. A weak-coupling BCS superconductivity is observed for all the compositions at $T_c$ = 1.43-1.65 K. It is found that 0.05 Ga atoms that rattle with a small Einstein temperature of $T_E$ = 8.1 K are introduced into the cage only for $x \geq 0.20$ in addition to ~0.3 Al rattlers with $T_E$ = 23 K. Correspondingly, $T_c$ jumps up by 8% at $x$ = 0.20, suggesting that low-energy Ga rattling enhances electron-phonon interactions. An electron-rattler interaction exists but is weak because of the nonionic nature of rattling Ga and Al atoms in the cage. In addition, the cage becomes soft for $x \geq 0.20$, which is also due to Ga rattling.

**Acknowledgments**

We are grateful to K. Hattori, T. Kobayashi, K. Ishida, M. M. Koza, and H. Mutka for helpful discussions. This work was supported by a Grant-in-Aid for Scientific Research on Priority Areas "Heavy Electrons" (No. 23102704) provided by MEXT, Japan.

1) B. C. Sales, D. Mandrus, and R. K. Williams: Science **272** (1996) 1325.
2) V. Keppens, D. Mandrus, B. C. Sales, B. C. Chakoumakos, P. Dai, R. Coldea, M. B. Maple, D. A. Gajewski, E. J. Freeman, and S. Bennington: Nature **395** (1998) 876.
3) Z. Hiroi, J. Yamaura, and K. Hattori: J. Phys. Soc. Jpn. **81** (2012) 011012.
4) T. Dahm and K. Ueda: Phys. Rev. Lett. **99** (2007) 187003.
5) K. Hattori and H. Tsunetsugu: Phys. Rev. B **81** (2010) 134503.
6) Z. Hiroi, S. Yonezawa, Y. Nagao, and J. Yamaura: Phys. Rev. B **76** (2007) 014523.
7) Y. Nagao, J. Yamaura, H. Ogusu, Y. Okamoto, and Z. Hiroi: J. Phys. Soc. Jpn. **78** (2009) 064702.
8) G. S. Nolas, J. L. Cohn, G. A. Slack, and S. B. Schujman: Appl. Phys. Lett. **73** (1998) 178.
9) S. Yonezawa, Y. Muraoka, Y. Matsushita, and Z. Hiroi: J. Phys. Condens. Matter **16** (2004) L9.
10) M. Brühwiler, S. M. Kazakov, J. Karpinski, and B. Batlogg: Phys. Rev. B **73** (2006) 094518.
11) T. Terashima, S. Uji, Y. Nagao, J. Yamaura, Z. Hiroi, and H. Harima: Phys. Rev. B **77** (2008) 064509.
12) T. Terashima, N. Kurita, A. Kiswandhi, E.-S. Choi, J. S. Brooks, K. Sato, J. Yamaura, Z. Hiroi, H. Harima, and S. Uji: Phys. Rev. B **85** (2012) 180503.
13) P. J. Brown: Acta Cryst. **10** (1957) 133.
14) A. E. Ray and J. F. Smith: Acta Cryst. **10** (1957) 604.
15) A. D. Caplin and L. K. Nicholson: J. Phys. F **8** (1978) 51.
16) A. Onosaka, Y. Okamoto, J. Yamaura, and Z. Hiroi: J. Phys. Soc. Jpn. **81** (2012) 023703.
17) A. D. Caplin, G. Grüner, and J. B. Dunlop: Phys. Rev. Lett. **30** (1973) 1138.
18) J. R. Cooper: Phys. Rev. B **9** (1974) 2778.
19) G. J. Legg and P. C. Lanchester: J. Phys. F: Met. Phys. **8** (1978) 2125.
20) T. Claeson and J. Ivarsson: Commun. Phys. **2** (1977) 53.
21) D. J. Safarik, T. Klimczuk, A. Llobet, D. D. Byler, J. C. Lashley, J. R. O'Brien, and N. R. Dilley: Phys. Rev. B **85** (2012) 014103.
22) M. Tinkham: *Introduction to Superconductivity* (Dover Publications, Inc., New York, 1996) p. 135.
23) N. E. Phillips: Phys. Rev. **114** (1959) 676.
24) L. R. Testardi: Rev. Mod. Phys. **47** (1975) 637.
25) S. D. Bader, G. S. Knapp, S. K. Sinha, P. Schweiss, and B. Renker: Phys. Rev. Lett. **37** (1976) 344.
26) G. Chanin, E. A. Lynton, and B. Serin: Phys. Rev. **114** (1959) 719.
27) A. Kontio and E. D. Stevens: Acta Cryst. A **38** (1982) 623.
28) B. C. Sales, B. C. Chakoumakos, R. Jin, J. R. Thompson, and D. Mandrus: Phys. Rev. B **63** (2001) 245113.
29) K. Kaneko, N. Metoki, H. Kimura, Y. Noda, T. D. Matsuda, and M. Kohgi: J. Phys. Soc. Jpn. **78** (2009) 074710.
30) J. Kunes, T. Jeong, and W. E. Pickett: Phys. Rev. B **70** (2004) 174510.
31) J. Yamaura, S. Yonezawa, Y. Muraoka, and Z. Hiroi: J. Solid State Chem. **179** (2006) 336.
32) K. Matsuhira, C. Sekine, M. Wakeshima, Y. Hinatsu, T. Namiki, K. Takeda, I. Shirotani, H. Sugawara, D. Kikuchi, and H. Sato: J. Phys. Soc. Jpn. **78** (2009) 124601.

33) J. Yamaura and Z. Hiroi: J. Phys. Soc. Jpn. **80** (2011) 054601.
34) T. Hasegawa, Y. Takasu, N. Ogita, M. Udagawa, J. Yamaura, Y. Nagao, and Z. Hiroi: Phys. Rev. B **77** (2008) 064303.
35) R. Saniz, J. E. Medvedeva, L. H. Ye, T. Shishidou, and A. J. Freeman: Phys. Rev. B **70** (2004) 100505(R).
36) Z. Hiroi, J. Yamaura, S. Yonezawa, and H. Harima: Physica C **460-462** (2007) 20.

Table I. Superconducting, electronic, and lattice properties of $A_xV_2Al_{20}$: $A_x = Al_{0.3}$ and $Ga_x$ with $x = 0.05-0.60$

| $x$ | ($Al_{0.3}$) | 0.05 | 0.10 | 0.15 | 0.20 | 0.30 | 0.40 | 0.50 | 0.60 |
|---|---|---|---|---|---|---|---|---|---|
| $a$ (nm) | 1.45157(8) | 1.451179(1) | 1.451214(1) | 1.451445(1) | 1.452034(1) | 1.452023(1) | 1.452452(1) | 1.419616(1) | 1.425052(3) |
| $\delta$ (Al) | 0.28(1) | 0.26(1) | 0.30(1) | 0.29(1) | 0.25(1) | 0.26(1) | 0.25(1) | 0.33(1) | 0.35(1) |
| $T_E$ (Al) (K) | 23.7(2) | 23.3(1) | 24.2(1) | 24.0(1) | 23.4(3) | 24.8(4) | 24.2(4) | 26.0(4) | 27.8(5) |
| $\delta$ (Ga) | | | | | 0.050(2) | 0.06(1) | 0.044(2) | 0.055(3) | 0.055(3) |
| $T_E$ (Ga) (K) | | | | | 8.1(1) | 8.0(1) | 7.2(2) | 7.4(1) | 8.0(1) |
| $T_D$ (K) | 420 | | | | 96 | | 92 | 86 | 88 |
| $T_c$ (K) | 1.49 | 1.53 | 1.53 | 1.53 | 1.65 | 1.60 | 1.54 | 1.50 | 1.43 |
| $\Delta C/\gamma T_c$ | 1.40 | 1.20 | 1.35 | 1.35 | 1.35 | | 1.40 | 1.45 | 1.35 |
| $2\Delta/k_BT_c$ | 2.7 | 2.6 | 2.9 | 2.8 | 2.9 | | 2.5 | 3.3 | 2.7 |
| $B_{c2}(0)$ (mT) | | | 65 | | 109 | | | | 137 |
| $\gamma$ (mJ K$^{-2}$ mol$^{-1}$) | 33 | 34 | 35 | | 35 | | 39 | 38 | 40 |
| $\chi_p$ [10 K] (10$^{-4}$ cm$^3$ mol$^{-1}$) | 6.61 | | 7.19 | | 6.00 | 6.04 | 5.71 | 5.26 | 5.01 |
| $R_W$ | 1.32 | | 1.36 | | 1.11 | | 0.94 | 0.88 | 0.78 |

13